\documentclass[sigconf]{acmart}
\usepackage{booktabs} 
\usepackage{bm}
\usepackage{color}
\usepackage{amsmath}
\usepackage{graphicx}
\usepackage{tabularx}
\usepackage{xspace}
\usepackage{algorithm}
\usepackage[noend]{algpseudocode}
\usepackage{caption}
\usepackage{subcaption}
\usepackage[flushleft]{threeparttable}

\algrenewcommand\algorithmicforall{\textbf{foreach}}
\algrenewcommand\algorithmicindent{.8em}
\AtBeginDocument{%
  \providecommand\BibTeX{{%
    \normalfont B\kern-0.5em{\scshape i\kern-0.25em b}\kern-0.8em\TeX}}}

\setcopyright{acmcopyright}
\copyrightyear{2021}
\acmYear{2021}
\acmDOI{10.1145/1122445.1122456}

\acmConference[WWW KMECommerce 2021]{WWW KMECommerce 2021: ACM Knowledge Management in e-Commerce Workshop}{}{}

\begin{document}

\title{Grouping Search Results with Product Graphs in E-commerce Platforms}

\author{Suhas Ranganath}
\affiliation{%
  \institution{Walmart Labs}
  \city{Bangalore}
  \country{India}
  \postcode{560103}
}
\email{Suhas.Ranganath@walmartlabs.com}

\author{Shibsankar Das}
\affiliation{%
  \institution{Walmart Labs}
  \city{Bangalore}
  \country{India}
  \postcode{560103}
}
\email{Shibsankar.Das@walmart.com}

\author{Sanjay Thillaivasan}
\affiliation{%
  \institution{Walmart Labs}
  \city{Bangalore}
  \country{India}
  \postcode{560103}
}
\email{Sanjay.Thillaivasan@walmartlabs.com}

\author{Shipra Agarwal}
\affiliation{%
  \institution{Walmart Labs}
  \city{Bangalore}
  \country{India}
  \postcode{560103}
}
\email{Shipra.Agarwal@walmartlabs.com}

\author{Varun Srivastava}
\affiliation{%
  \institution{Walmart Labs}
  \city{Bangalore}
  \country{India}
  \postcode{560103}
}
\email{Varun.Srivastava0@walmartlabs.com}

\renewcommand{\shortauthors}{}


\begin{abstract}
Showing relevant search results to the user is the primary challenge for any search system. Walmart e-commerce provides an omnichannel search platform to its customers to search from millions of products. This search platform takes a textual query as input and shows relevant items from the catalog. One of the primary challenges is that this queries are complex to understand as it contains multiple intent in many cases. This paper proposes a framework to group search results into multiple ranked lists intending to provide better user intent. The framework is to create a product graph having relations between product entities and utilize it to group search results into a series of stacks where each stack provides a group of items based on a precise intent. As an example, for a query "milk," the results can be grouped into multiple stacks of "white milk", "low-fat milk", "almond milk", "flavored milk". We measure the impact of our algorithm by evaluating how it improves the user experience both in terms of search quality relevance and user behavioral signals like \textbf{A}dd-\textbf{T}o-\textbf{C}art.
\end{abstract}






\maketitle

\section{Introduction}

\begin{figure*}
\begin{subfigure}[b]{0.6\textwidth}
\includegraphics[width=\linewidth]{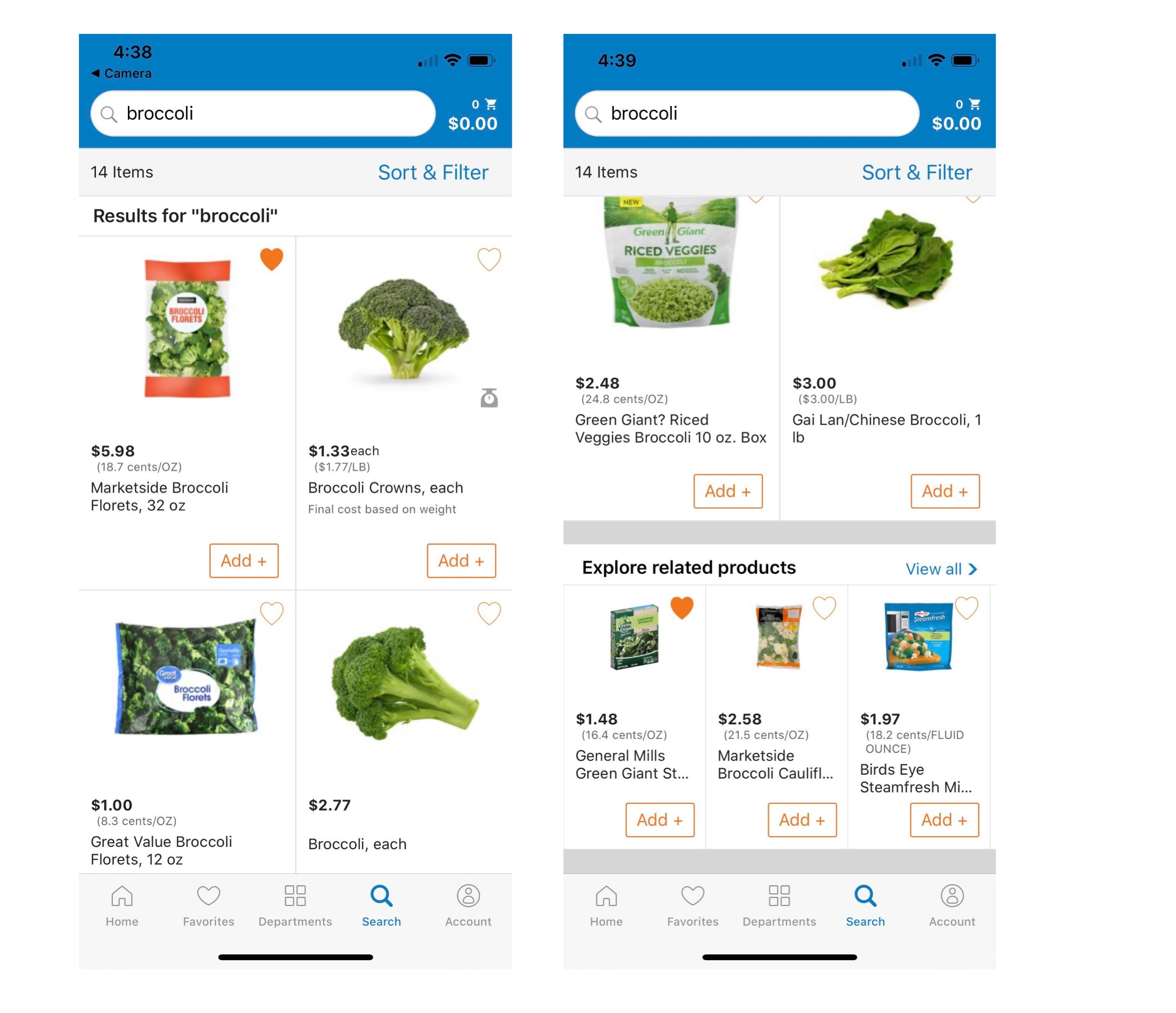}
\caption{Grouping search results for "broccoli"}
\label{fig:intro}
\end{subfigure}%
\begin{subfigure}[b]{0.5\textwidth}
\includegraphics[width=\linewidth]{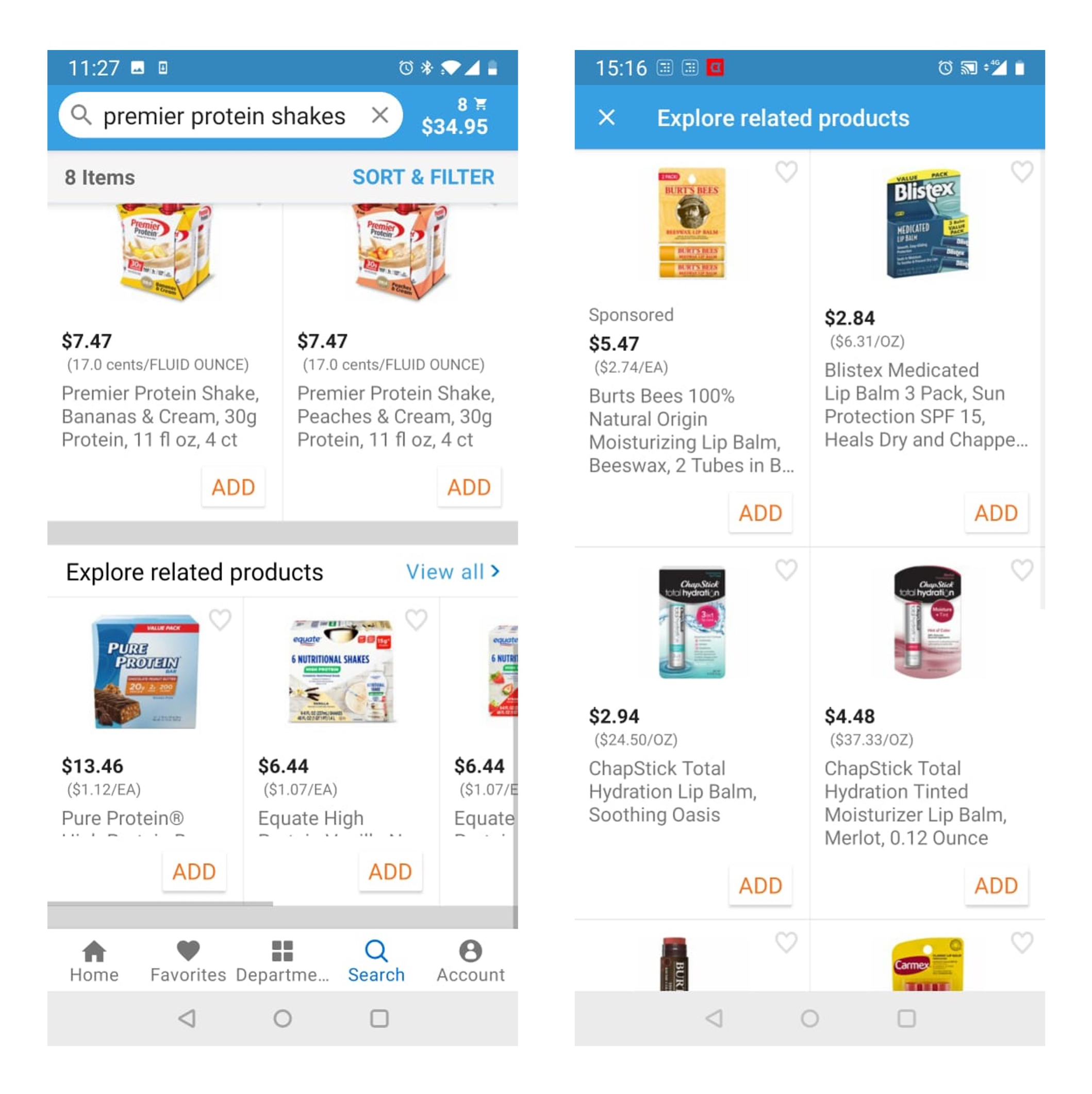}
\caption{Grouping search results for "premier protein shakes"}
\label{fig:intro2}
\end{subfigure}%
\caption{Example of Grouping Search Results on Walmart App}\label{fig:stacked_recall}
\end{figure*}

One of the most common problems in any search platform is understanding the user intent as specified in the textual query and providing search results based on that. Understanding user intent is sometimes very challenging because the intent can be unclear and can have multiple meanings. It may be challenging to satisfy users with a single ranked list representing all possible intents in such cases. 

In this paper, we propose a framework to solve this problem in an unorthodox way. We have grouped search results into multiple ranked lists of items based on a logical group so that for each stack, users find an exact search result. For a given query, we are stacking multiple ranked lists where each group comprises of items based on a logical intent. There could be several dimensions in which search results can be grouped. A pair of stacks could be comprised of similar products or substitute products or complementary products. A few examples illustrating the concept is presented in Fig \ref{fig:stacked_recall}, and we next describe them. 

The first example illustrated in Fig \ref{fig:stacked_recall} of the match is based on perfect vs. approximate match items. On the left, we can see a single ranked list containing different possible items with the exact intent "broccoli" and "broccoli mixed with other vegetables". In the proposed stacked recall experience, the search platform shows two stacks of results; the primary stack contains results based on a perfect match for the intent expressed in the query term. The secondary stack includes results based on an approximate match. Similarly, for the query "oranges", in the vanilla version of the search, we show "Orange Fruits", "Orange Flavoured Juices", "Orange Drinks" all mixed in one group. In the new experience, we propose to show the perfect matches of "Orange Fruits" all grouped in the primary stack, followed by "Orange" flavored items in the secondary stack and potentially other "Fruits" from the user's purchase history in the third stack. 

The next example illustrated in Fig \ref{fig:stacked_recall} is based on perfect items and similar items. This use-case primarily applies to queries where there is less assortment to satisfy the exact intent. The proposed experience can show the matching results for the query, and the secondary stack shows products similar to the query intent. So, the secondary stack has the opportunity to recommend similar items. For example, a query "premier protein shake" has an exact intent, and the number of products relevant to the search query is limited. In the proposed stacked recall experience, we show all "protein shake" of the brand "premier" in the primary stack for queries with a minimal number of items in the resultant set. For the second stack, the framework recommends items from different brands of protein shakes and product types with intents similar to protein shakes. Similarly, for a query "Starbucks coffee beans," the primary stack will show "coffee beans" from "Starbucks" whereas the secondary stack will show coffee beans from "Lavazza".

We next survey literature related to our work and put it in context with different literature.

\section{Previous Work}
Product recommendation in e-commerce search has been one of the most attractive fields of research for more than a decade. Traditional researches used Collaborative Filtering, Non-Negative Matrix Factorization \cite{lee2001algorithms}, \cite{gouvert2020ordinal} based methods to recommend relevant content. Tapestry \cite{goldberg1992using} was one of the earliest implementations of collaborative filtering. Afterward, there were plenty of researches on using Collaborative Filtering \cite{melville2002content}, \cite{schafer2007collaborative}, \cite{sarwar2001item}, \cite{das2007google}, \cite{linden2003amazon}, \cite{su2009survey} in e-Commerce product recommendation, song or movie recommendation, etc.  In recent times, researchers have proposed Neural network based collaborative filtering \cite{10.1145/3038912.3052569}.

In recent times, there is a lot of focus on learning low-dimensional representation of entities using Graph structured data in the e-commerce domain. Graph Representation Learning aims to learn a low-dimensional representation of nodes through neighborhood aggregation. Node2vec \cite{10.1145/2939672.2939754}, DeepWalk \cite{10.1145/2623330.2623732}, LINE \cite{10.1145/2736277.2741093} were the pioneer in learning graph representation which was built on the fundamentals of a random walk. GraphSAGE \cite{hamilton2017representation}, \cite{pmlr-v80-xu18c} used various methods to aggregate information from neighbors and learn representation accordingly. Whereas the Graph Convolution Networks \cite{kipf2016semi},\cite{zhang2019star} tries to aggregate information from immediate neighbors using convolution approaches.

Taxonomies are widely used in Online Retail for product recommendation or in web search engines to enhance query understanding. In \cite{shen2020taxoexpan}, Shen et al. proposed learning taxonomy using graph neural network. 
The other set of works are on the fundamentals of a Graph-based Recommendation system. In this set of work, user engagement is looked at as a bipartite graph. From the user's engagement, similar products are computed based on co-purchased items. In \cite{berg2017graph}, Berg et al. proposed a Graph Auto-Encoder to complete the bipartite graph, which can be used to predict links in the bipartite graph. In \cite{10.1145/3219819.3219890}, Rex et al. proposed a data-efficient Graph Convolution Networks algorithms called PinSage, which combines efficient random walk and graph convolution to learn a representation of nodes. \cite{cen2019representation},\cite{zheng2019robust}, \cite{fu2020fairness} are the other set of work on representation learning using Graph.

Grouping Web image search was proposed in \cite{10.1145/1027527.1027632}. This was the first of a kind method to group search results. In \cite{10.1007/978-3-540-72524-4_70} Reza et al. proposed a method to group search results based on a different meaning of the query. It uses WordNet to determine the meaning of search items. This work focuses on grouping search results based on the user's broad intent specified in the query. There are a few more research \cite{zhang2020stacked},\cite{zhang2020cross},\cite{kuang2019fashion},\cite{xu2019relation} on e-commerce domain where product recommendation was learned using Graph representation which were built using user-item engagement data.
\begin{figure*}[ht]
    \centering
    \includegraphics[scale=0.3]{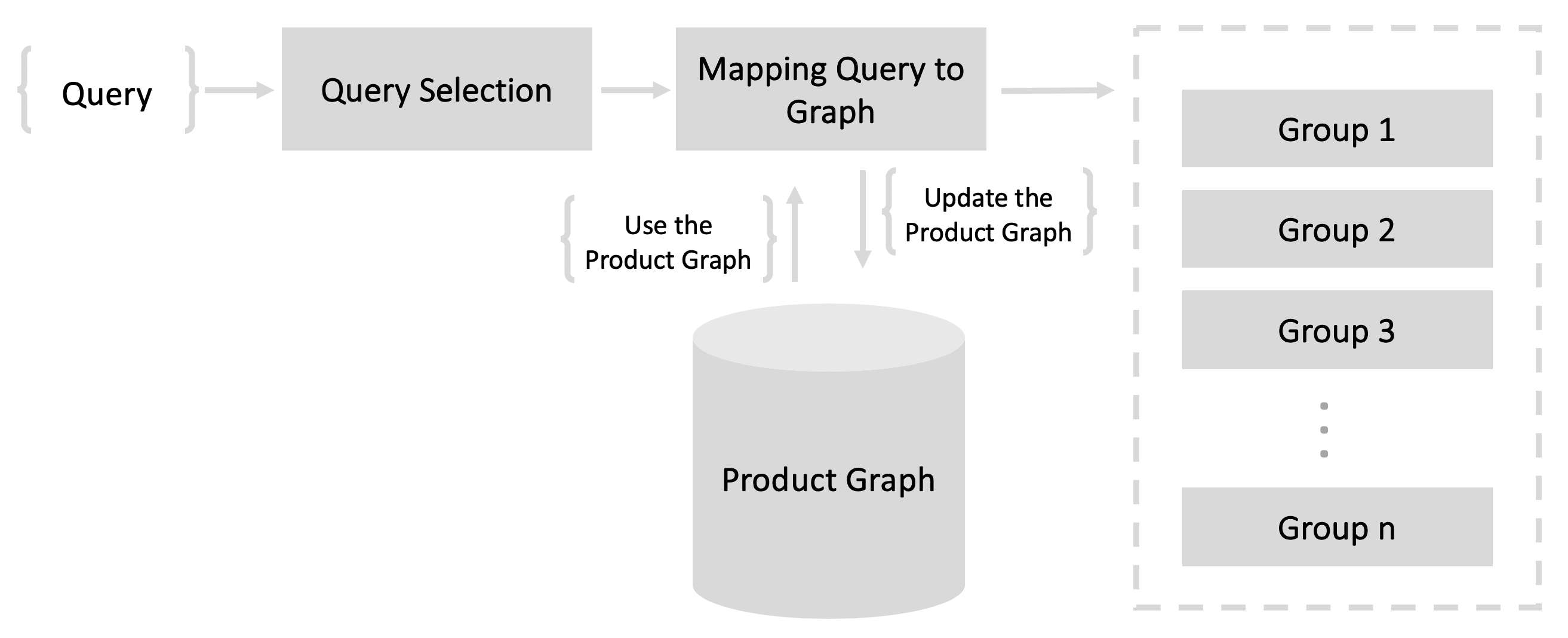}
    \caption{Framework to group queries using Product Graph}
    \label{fig:stacked_recall_flow}
\end{figure*}

\section{Methodology}
Next, We will describe our framework to group search results for e-commerce queries. The framework is illustrated in Fig \ref{fig:stacked_recall_flow}. A set of criteria were developed to decide whether a query is eligible for "Stacked Recall" experience. So, first of all, a query is validated for it's eligibility for Stacked Recall. We then map the queries to attributes in the Product Graph, which is an internally developed, periodically updated set of relations between product attributes. The subsequent section will describe how we use the mapped attributes and their relations to group search results into a series of stacks.

\subsection{Selection of Eligible queries}
To optimize user experience, we have developed criteria to identify queries to show stacked recall. We break the query into a set of intents corresponding to the different attributes in the catalog. The most commonly seen intent in a user query are Product Type - which denotes the type of product that the users want to buy and Brand - which denotes the brand of the product that the user intends to buy. 

Let an intent $i$ be denoted as a set of attributes $i=\{a_1,a_2,a_3\}$, and the set of intents be denoted by $i \in \mathcal{I}$ and the corresponding orders be denoted by $\mathbf{o}_i$ for all $i \in \mathcal{I}$. Let the set of features related to each intent $f_i $ $\forall i \in \mathcal{I}$. We then construct a threshold algorithm across the feature set. We construct the eligible set $i \in \mathcal{E}$ by selecting all the intents $i$ whose feature values is lower than the selected threshold. We say a given query is eligible if it has an intent which can belong to the eligible set $i \in \mathcal{E}$.
 \begin{figure}[h]
\caption{Engagement loss distribution}
\centering
\includegraphics[width=0.5\textwidth]{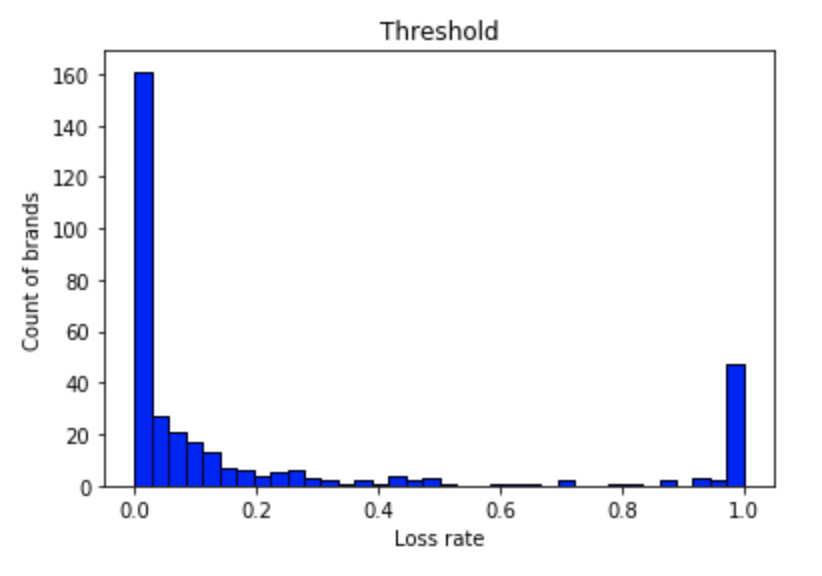}
\label{fig:engagement_threshold}
\end{figure}
Next, we describe two of the features that we are using to illustrate the criterion to select a query to optimize the user experience. The first feature is the engagement loss of the query. Whether or not to show a grouped rank list is based on whether the user generally honors the intent for the query. We pick queries where grouping search results according to the user intent result in a minimal loss in user engagement.
\begin{equation}
    e_{i} = \frac{\sum_{\forall{i \not \in \mathcal{I}}}{\mathbf{o}_i}}{\sum_{i \in \mathcal{I}}{\mathbf{o}}}
\label{eqn:loss}
\end{equation}
The user behavior has localized effects and we compute the engagement loss for all intents. We compute the engagement loss for each set of intents that we have seen being expressed in the query as seen in Equation \ref{eqn:loss}. If the query's intent belongs to a large number of items, the diversity of the results can be lost if the number of items is large, leading to a loss in engagement. To counter this, another example feature set is 
\begin{equation}
    count_{i} = \cap{a  \in i}{\mathbf{c}_a}
\label{eqn:loss2}
\end{equation}
The data collection takes the last six months' queries with orders greater than an order threshold as the valid queries. We understand the intent of the eligible user queries and then generate the set of intents $\mathcal{I}$. We then apply to preprocess like stemming and stop-word removal on the queries and items. We finally compute the feature set like engagement loss and item counts for each query intents set. For the set of features, we plot the distributions of these features, similar to Fig \ref{fig:engagement_threshold}, and use internal threshold algorithms and pick a set of items whose corresponding feature sets which is greater than the threshold. 

Next, we map the eligible queries to the attributes in the product graph. We first describe the creation of the product graph and then illustrate how we map the queries to the graph's attributes.

\begin{figure*}
\begin{subfigure}[b]{0.5\textwidth}
\includegraphics[width=\linewidth]{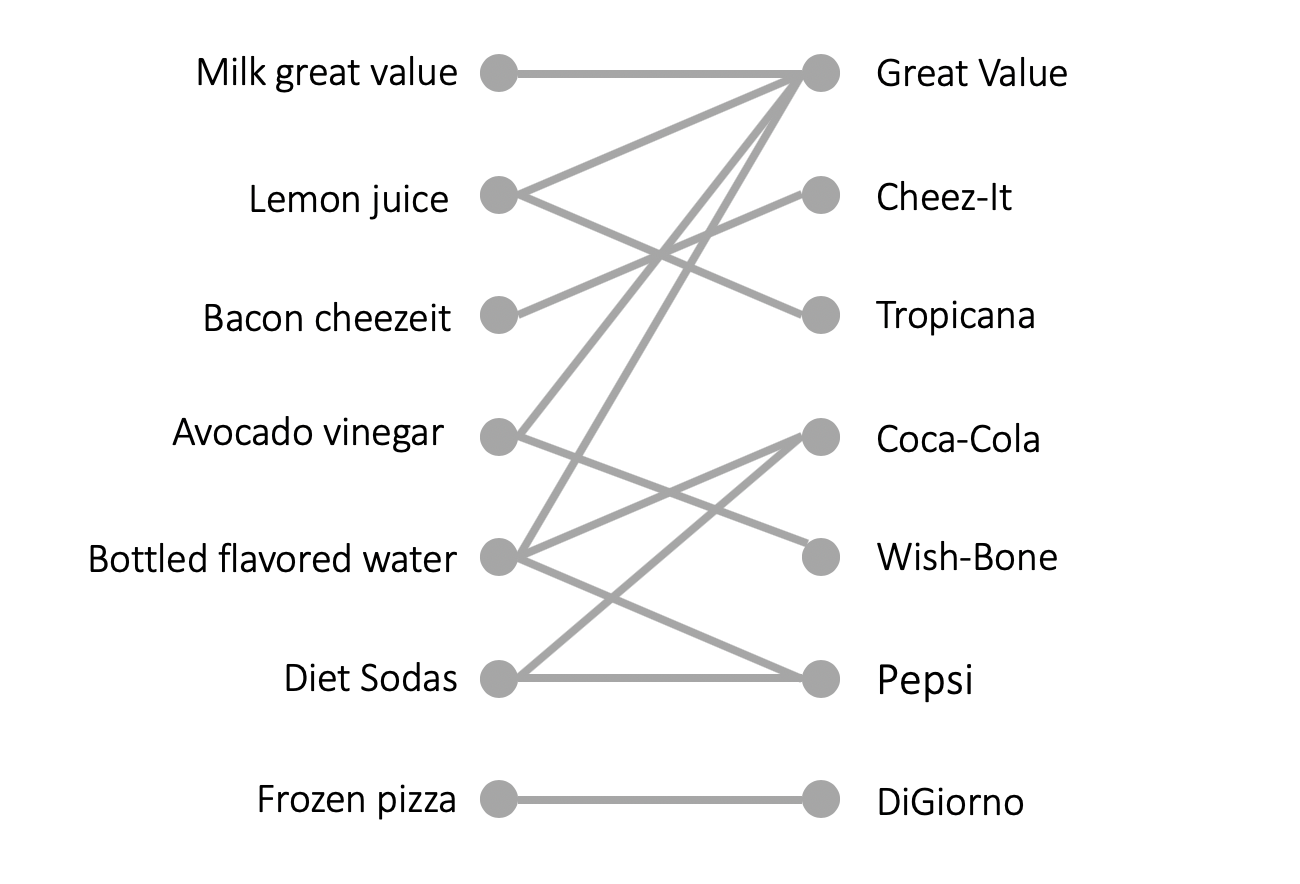}
\caption{Query attribute engagements}
\label{fig:gull}
\end{subfigure}%
\begin{subfigure}[b]{0.6\textwidth}
\includegraphics[width=\linewidth]{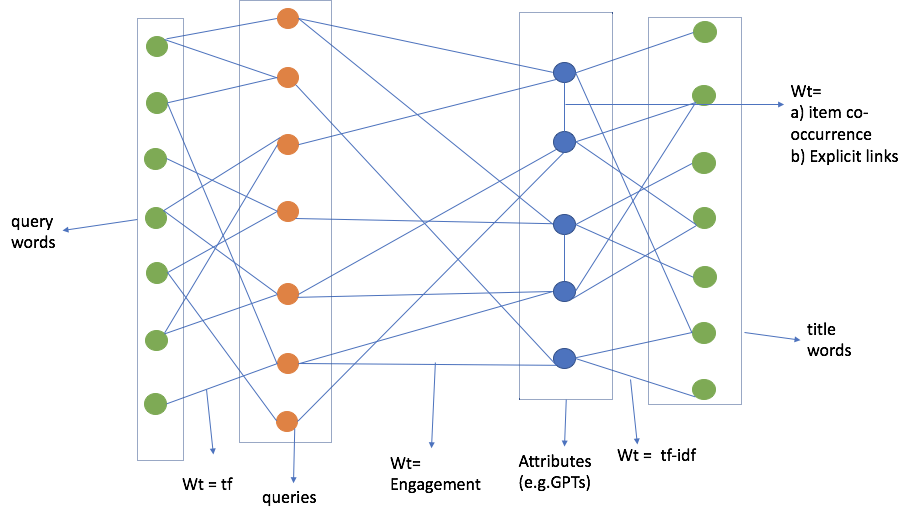}
\caption{Full query attribute product graph}
\label{fig:gull2}
\end{subfigure}%
\caption{Product graph construction}\label{fig:animals}
\end{figure*}

\begin{figure*}[ht] 
    \centering
    \includegraphics[scale=0.35]{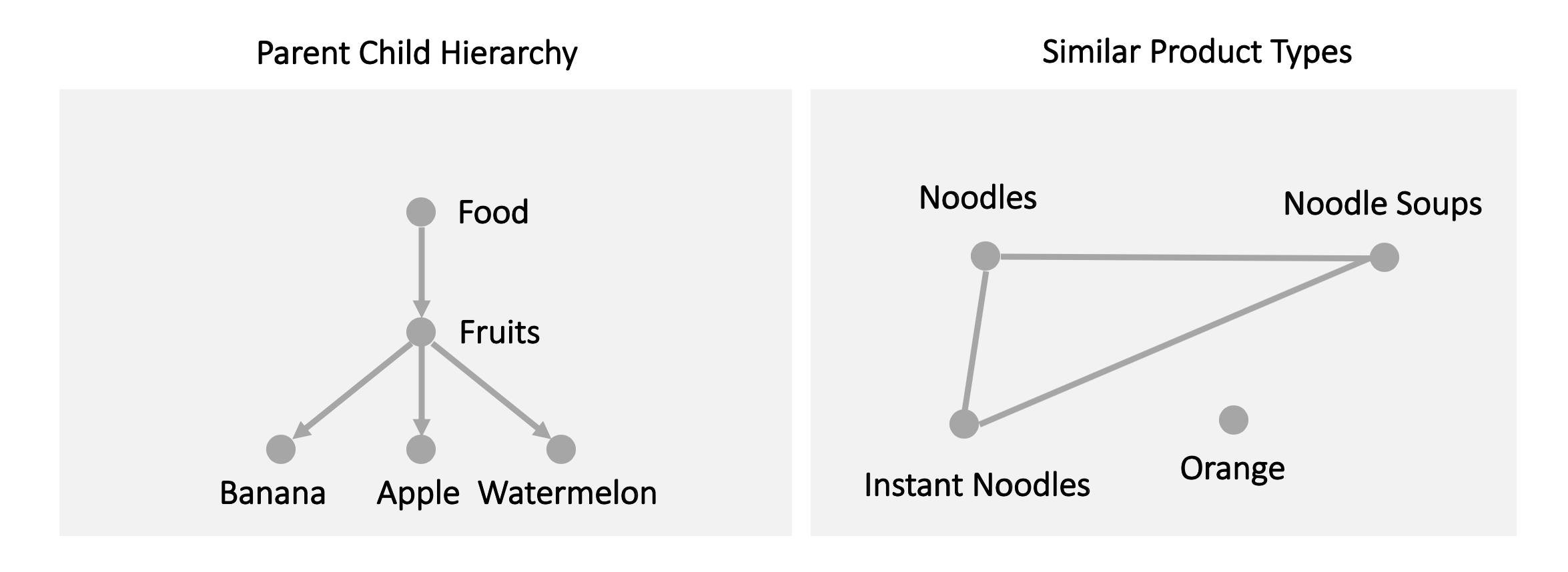}
    \caption{Relationship between e-commerce queries, products and attributes}
    \label{fig:ecommerce_graph}
\end{figure*}

\subsection{Product Graph Creation} \label{RetailGraphCreation}

The Product Graph is used to construct relations between attributes of products. We utilize the Product Graph to facilitate the grouping of products appearing in the search result. Primarily, two kinds of relations exist in the Product Graph: Parent-Child relation and Lateral relation. The relations can be either created manually by expert taxonomists or inferred using algorithms. The different kinds of relations are illustrated in Fig \ref{fig:ecommerce_graph}. 
 
We first examine how the hierarchical structure for a given query is formed. The hierarchical product tree organizes the individual products into a tree. In Fig \ref{fig:ecommerce_graph}, we can see the hierarchical Product Type structure of fruit-related Product Types.
The hierarchical structure is mainly conceptual and hence created by professional taxonomists. For a specific query "apple," we have to traverse the tree to get to the "Apple" node. At the first level, we have to identify "Banana" belongs to a Product Type as "Fruit" and then identify among its children the Product Type of "Apple." The next kinds of relations in the Product Graph are the lateral relations between attributes. These relations are inferred based on large scale interactions of the users with different products, and hence tagging by expert taxonomists may not be scalable. In Fig \ref{fig:ecommerce_graph}, we see that relations between "Noodles," "Noodle Soups," and "Instant Noodles" in that users interact with these products in a nearly identical manner.

We next describe the process of creating different kinds of Product Graphs in detail. We collect data from the interactions and build a set of relational features that we use to construct a Product Graph. As Users interact with the Walmart e-commerce platform by performing actions like searching for an item, typing query, click on product details, Add items To Cart (called as ATC), and purchasing items, the platform generates signals based on these events. These signals are used to improve relevance for user queries during run time. We collect the user query, the items they have searched and purchased over the last six months, and the attributes such as Brand, Product Type, etc. of the engaged items. We then perform a series of pre-processing steps to obtain useful information out of the query item pairs, which has at least 10 Add To Cart items and ten orders. Using all of these processes, over 2 million data was generated through the feature engineering process. We now describe the way we use the data as per Table \ref{table:dataset} to infer edges between attributes.
\\
 \begin{table}[ht]
  \centering
  \vskip-15pt
  \caption{Statistics of the Dataset} 
  \label{table:dataset}
  \begin{tabular}[t]{lr}
    \toprule
     \textbf{Entities} & \textbf{Statistics} \\
    \midrule
    Unique Queries & 816452  \\
    Items &  90658 \\
    Attributes - Product Types &  89274 \\
    Attributes - Global Product Types &  2501 \\
    Attributes - Brands & 9207  \\
    Add to Carts & 10\\
    Number of nodes & 816452 \\
    Number of edges & 2143444 \\
    \bottomrule
  \end{tabular}
\end{table}

We primarily group items based on their attributes. To achieve this, we first construct a graph to infer the relations between the attributes. Broadly speaking, we deem two attributes to be similar if similar queries engage them. We further divided into two approaches. The attributes can be of bilateral like two Product Types, or multilateral like Brand - Product Type - Brand, which gives Brand - Brand relations through different Product Types. Different graph inference techniques can be designed to infer these relations; we describe two of the techniques applied.

After formulating the initial graph based on the user-item engagement data as described earlier, we use Graph Neural Network-based algorithm to learn a low-dimensional representation of nodes called embedding. Graph embedding helps us learn relationships in the graph-structured data and reduce noise and infer new links efficiently. In this paper, we learn the latent space representation of nodes in an unsupervised way.

We created a series of bipartite graphs with queries, query terms, item titles, and attributes. This graph is illustrated in Figure \ref{fig:gull2}. To construct this, we aggregate items based on their attribute values such as Brand, Product Type. We deem queries, query terms, attributes, and attribute terms as vertices. As seen in Figure \ref{fig:gull2}, we have three kinds of edges: query terms - query, query - attributes, and attribute terms - attributes. The edge weights between query terms - query and attribute - the term frequency defines attribute terms. The query's engagement gives the edge weights between query and attributes to the attributes.  When a new query appears, we extract intents of the query using a model named "Query Understanding". For the interest of this paper, "Query Understanding" algorithm was not illustrated. We will use this graph to find groups of attributes that are similar and make an additional baseline.

We have implemented Variational Graph Auto-Encoder (VGAE)\cite{kipf2016variational} to learn low-dimensional representation in an unsupervised way. We first encode the graph into embedded vectors by using a Graph Convolution Network (GCN) \cite{kipf2016semi} - based encoder and a simple inner product based decoder. The encoder learns a transformation from the initial node representation to a latent space representation that understand the context of product title. The decoder does an dot product on the latent space representation to reformulate a complete graph. During training, the model tries to minimize the reconstruction loss between the initial graph and the graph approximated by the decoder. 

We have trained a fasttext \cite{bojanowski-etal-2017-enriching} unsupervised model to learn initial low-dimensional representation of nodes summarized in a $N*D$ matrix $X$ where $N$ is the number of node in the Graph and $D$ is the dimension of initial representation of the individual node generated using fasttext \cite{bojanowski-etal-2017-enriching} model.
So, the encoder in \cite{kipf2016variational} takes as input an undirected graph $G=(V,E)$ in the form of adjacency matrix $A$ and the initial node representation $X$. 

Let's introduce a latent variable $Z_i$. The latent space representation from the Encoder output is summarized in a $N*F$ matrix $Z$ where $N$ is the number of nodes and $F$ is the dimension of latent space representation. We implemented the method mentioned in \cite{kipf2016variational} where the decoder is parameterized by a 2 layer GCN \cite{kipf2016semi}.
\begin{equation}
GCN(X,A) = \Tilde{A}ReLU(AXW_0)W_1
\end{equation}
where $\Tilde{A}$ is the normalized adjacency matrix. The adjacency matrix is normalized by the degree of nodes. The encoder learns the latent space representation $Z$ and the decoder tries to reconstruct the adjacency matrix $\Tilde{A}$ as follows:
\begin{equation}
 \Tilde{A} = \sigma(ZZ^T)    
\end{equation}
      
where $Z = GCN(X,A)$



\subsection{Mapping queries to product graph}
We first describe how we traverse the product type hierarchy for a given query. After mapping the user intent to appropriate nodes in the hierarchy, we utilize the attributes and their relations in the product graph to group products appearing in the search results.
\subsubsection{Mapping queries to attribute hierarchy}
Once we built the product type hierarchy, the next task is to map a given query to the hierarchy. We build an algorithm and learn to map queries to traverse the category tree. We train a convolution neural network-based classifier for each level of the hierarchy. The training framework has the advantages of scalability and making it faster to serve the millions of online users. 
 
The training data for each level is based on the past engagement of the query for a product type in that level and the semantic similarity between the query and items in the product type of that level. For a new query, we first map the query to product types at each level. After that, we perform a top-down traversal of the hierarchy to get the final output. The design of the hierarchical classifier illustrated in Fig 2. In the next paragraph, we will describe each of these phases. As shown in the figure, each classifier per level has three stages; the initial feature embedding, convolutional neural network layer, and a weighted softmax layer in the end. 
 
Let the set of queries for training data be given by  $Q \in \mathcal{R}^{q X t}$, where $q$ is the number of queries, and $t$ is the number of unique terms in the query set. We train a skip-gram word2vec model on past 6 months queries and product titles from the item catalogue to get embeddings for each term $V \in \mathcal{R}^{t x n}$ where n is the number of dimensions of the embeddings. To derive initial embeddings for each query, we average out vectors of the corresponding terms to get $F \in \mathcal{R}^{q x n}$. In addition to this, we derive the target vector to denote the relationship of the query with the product type  $\textbf{e} \in \mathcal{R}^{1 X p}$ of the weight of query to product type $p$. Two metrics determine the target vector's values; the past engagement between the query and product type and the semantic similarity between query and items belonging to the product type.  
 
We have a convolutional network layer $CNN_l$ to map the query for each level in the hierarchy to obtain $x_l = CNN_l(F)$. For each layer, we have a softmax layer  $prob_{pl}=softmax(W * x_l')$, where $pl$ is the length of the product types in the level $l$. Each value in the $prob_{pl}$ gives the approximate probability of a query $q$ belonging to a product type. The true probability of query to the product type is given by the vector $e$, which we use as a target vector. We then maximize the likelihood of the target vector is close to the predicted vector $prob_l$ through the objective function $object = \max likelihood(e,p_l)$ and back-propagate it through the network to learn the network parameters for each layer.
 
When a new query comes in, we pass it through the learned network at each layer to obtain the probability of product types at each layer. We next need to obtain the product type hierarchy $h{}$ from the classifier's set to obtain the final product hierarchy from the classifier. We denote two attributes for each product type and $p.score$ initialized to zero, $p.children$ as the set of children of $p$ in the taxonomy, and $p.qc$ as the subset of $p.children$ which is relevant to the query. We traverse the classifiers at each level from the top one. At the first level, we select the set of product types $pt_1$ whose $p.score$ is above a threshold $t_1$ computed as a function of the score distribution. From the next level 2, we choose product type $pt_2$ which are both the above the threshold $t_2$, For each of the product types $pt$ in $pt_1$, we pick  $p.qc = pt_2 \cap p.children$, where $p.qc$ is the children of $p$ relevant to the query.

An additional problem in the traversal is to decide whether there is enough support to continue to the next level. For example, for a query "fruit" in Fig 1, we need to stop at level 2 in the product type "Fruits", failing which the accuracy may be affected. At each level $l$, we compute support $s_l$ use metrics based on engagement entropy and linguistic specificity. If the support $s_l$ is sufficiently high, we go to the next level, else we stop. In this manner, we output a product hierarchy for each given query. 


We conducted an offline and online evaluation of our mapping framework. For offline tests, internal subject matter experts evaluated the search results, and it gave a result of 1.5\% higher NDCG compared to production. For the online evaluations, we deployed the production model, and the visitors in the traffic segment serving the models added significantly more items to cart than the users in the control segment

\subsubsection{Consuming the Product Graph}
We next describe how we traverse the lateral relations after we map the query to the relational attributes. Before we proceed, we will try to define some basic terms to be on the same page. The ATC is added to cart action done by the user for a particular product. Substitute products are those on which the users have done an ATC action for the same query and which belong to a different category.  

We are using a baseline ranking algorithm as described for the given query for showing items of perfect and approximate in different stacks. This will provide results in the perfect and approximate items for user queries. Perfect vs. Approximate matches applicable to medium or narrow queries.  

Our baseline ranking can achieve fast search responses because, instead of searching the text directly, it searches an index instead. This type of index is called an inverted index because it inverts a page-centric data structure (document->item) to a keyword-centric data structure (item->document). On top of it, there are many features like past click engagement data, popularity, brand, and item sales, which act as signals. This layer will rerank the items which are returned from the baseline and in turn will be returned to the customer. The stacked recall algorithm will use this baseline layer to retrieve the item lists from the baseline.

\begin{figure*}[ht]
\caption{Grouping Search Results} 
\centering
\includegraphics[width=0.8\textwidth]{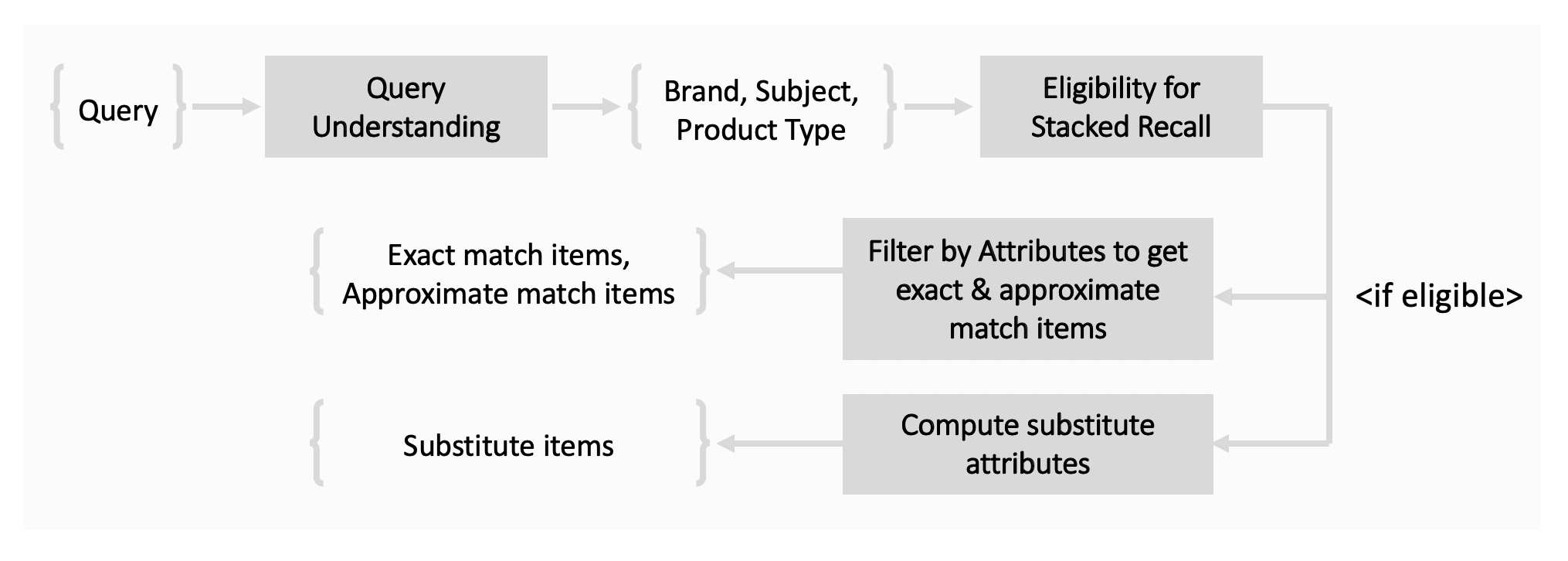}
\label{fig:groupresults}
\end{figure*}

There will be two baseline calls which will happen as per the figure \ref{fig:groupresults}to find the approx items/similar items and perfect items. The baseline will return 128 items per call, where it will be embedded in group results. There will be duplicates of the same items for two different baseline calls. We then use the duplicates algorithm, which is used to remove the duplicates of common items in both layers. e.g., "great value milk": predicted brand "Great Value," and the top predicted product type is "Milk" as per query understanding. We use the set of intents in the query to group items. For example, the brand "Great Value" and product type "Milk," the primary stack, will be correctly filled.

\section{Experimental results}
To validate the framework and showcase the framework's application, we conduct the following experiments; a)offline evaluations of the relevance of primary groupings vs. online b) online evaluations of groupings.

The following metrics were used to evaluate our algorithm
\begin{itemize}
    \item Normalized Discounted Cumulative Gain (NDCG): The expert annotators are given a ranked list for each query for the control without grouping and the primary set of the variation, and then they give a rating of 1 to 5, 5 being the best. We then compare the NDCG of the control and variation, $NDCG_k$ considers the order of the positive examples within the top K ranks. This measure is computed as 
    
        
    \begin{equation*}
    \begin{split}
        DCG_{k} &= \sum_{r=1}^{k}\frac{2^{sui(i)-1}}{\log(i+1)}\\
        nDCG_k &= \frac{DCG_k}{IDCG_k}
    \end{split}
\end{equation*}

where $IDCG_k$ computes the $DCG_k$ for the optimal ordering of candidate responders.
    \item Online performance latencies: Latencies measure the time it takes between when the query is issued and returned results. We then sort the traffic latencies in ascending order and obtain the times for different percentiles of the traffic. We measure the average, P@50, P@60, P@70, P@80, P@90, P@95, P@99, and P@99.9 denoting the $50^{th}$, $60^{th}$, $70^{th}$ percentile and so on.
    \item Online behavior metrics: In our online experiments, we expose one set of users to the production algorithm without grouping and another set of users to the grouped result set algorithm. We measure the average number of orders, conversions, and revenue for each query for the control and variation.
\end{itemize}
\begin{figure}[ht]
    \centering
    \includegraphics[scale=0.4]{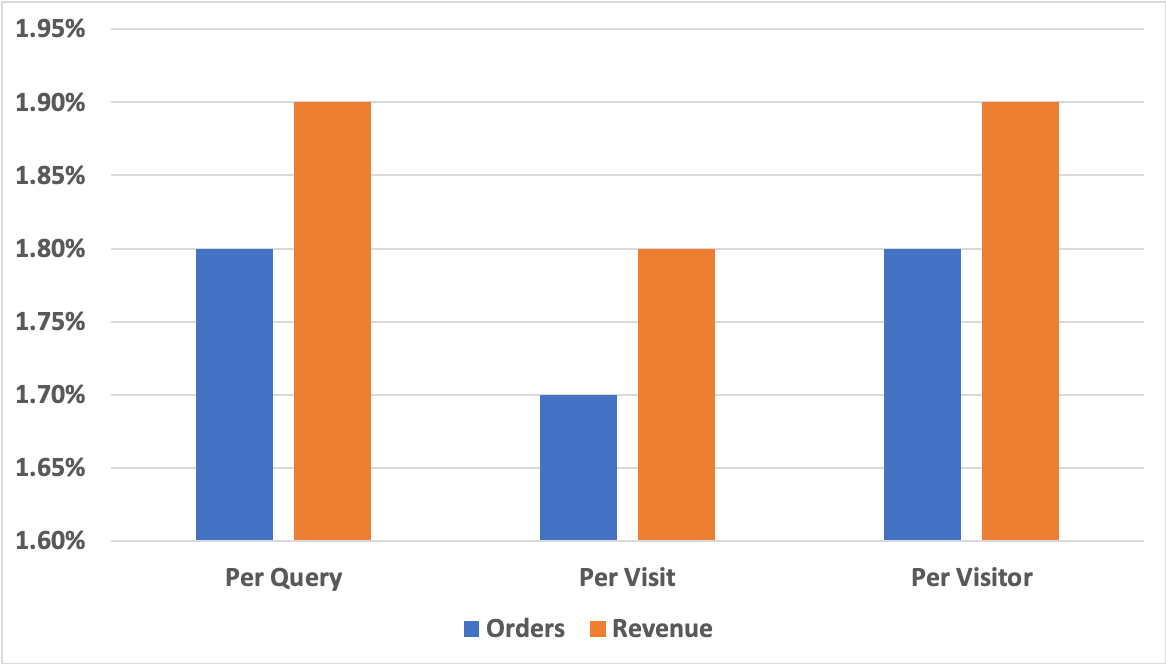}
    \caption{Metrics with online test. Grouping search results give a significant improvement in online metrics}
    \label{fig:online_metics}
\end{figure}

\begin{figure}[ht]
     \includegraphics[scale=0.4]{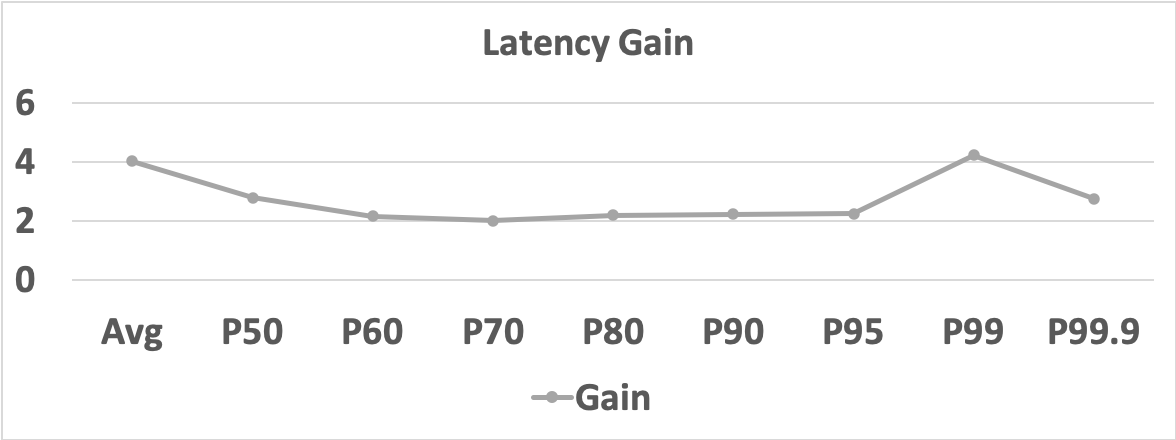}
    \caption{Percentage latency gain. Grouping search results give significant improvement in online metrics}
    \label{fig:latency_gain}
\end{figure}
    
\begin{figure}
     \includegraphics[scale=0.4]{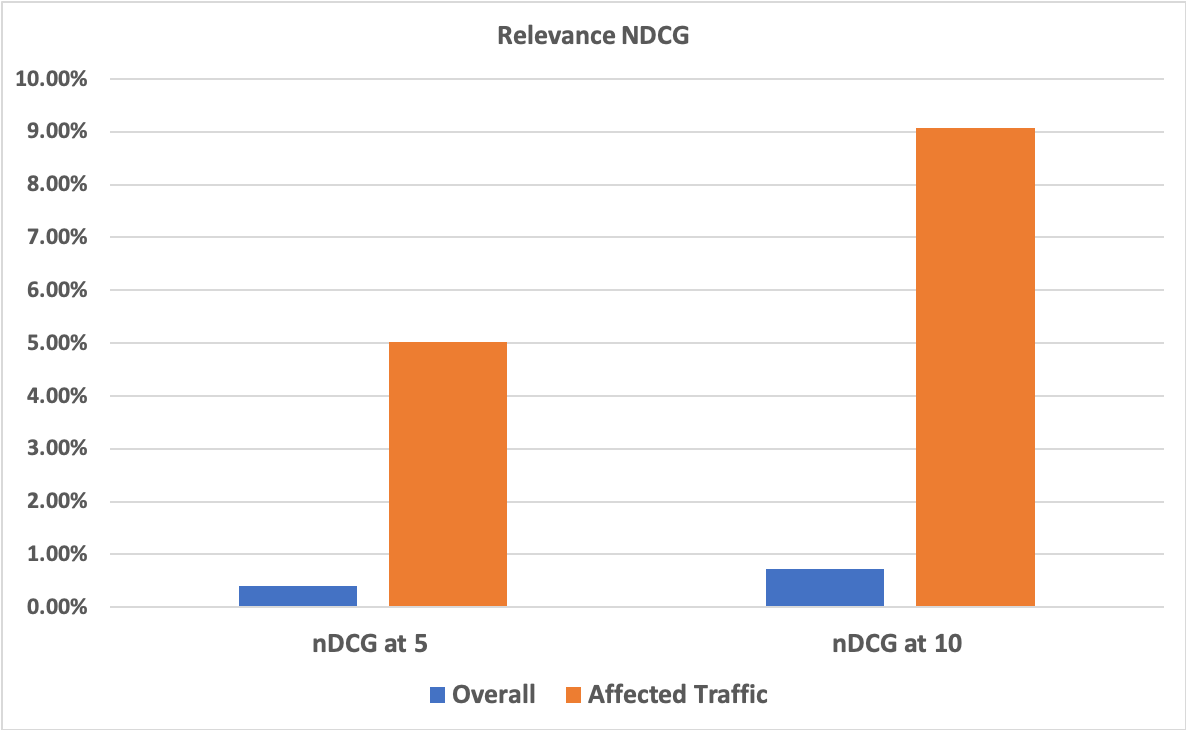}
    \caption{Percentage latency gain. Grouping search results give significant improvement in online metrics}
    \label{fig:relevance_ndcg}
\end{figure}

To evaluate our algorithm concerning the annotators, we give the control and variation ranked list to the expert annotators. The annotators are given a rating of 1 to 5, 5 being the best. The results of the relevant tests are illustrated in Fig \ref{fig:relevance_ndcg}. The algorithm affects 11.8\% of the overall traffic. From the figure, we can see that the large gains in both the metrics on the affected traffic show grouping search results help in showing relevant products to users. It also shows a positive gain in relevance across the overall traffic in both the metrics.
 
After verifying the algorithm through manual evaluation by experts, we deployed the models to production traffic. We first measure if the search results' grouping causes any performance issues and delay the response to the user. We measure the latency of the production and our framework for millions of queries and sort them. We obtain the return times of at different percentile of the traffic and compute the percentage difference between control and variation. We plot the difference in Fig \ref{fig:latency_gain}. From the figure, we can observe that the framework gives only a slight increase in latency over the production system. The framework will not affect the cost to serve the results to the user and showcases the efficiency of the system. 

Next, examine how users interact with the framework through an online AB test. We compare their behavior through average \textit{orders per query}, \textit{orders per visit}, \textit{orders per visitor}, \textit{revenue per query}, \textit{revenue per visit}, \textit{revenue per visitor} and plot it in Fig \ref{fig:online_metics}. From the figure, we see that the framework improves all online metrics, showcasing that grouping search results are received well by the users and can also increase the organizations' revenue and orders. This also showcases the framework's ability to group the search results in a way that can induce positive responses from millions of users.

\section{Conclusion and Future Works}
Walmart E-commerce platforms observe millions of users expressing shopping intents in varied ways. These intents are often not clearly expressed, and multiple intents are expressed through a single query. To respect such complex user intents, we describe the first of a kind framework in this paper to group e-commerce search results into multiple ranked lists. The framework maps the queries onto an internally constructed product attribute graph. Using the mapped attributes and their relations, we then group search results into multiple ranked lists. Our extensive offline and online testing shows that the framework minimally impacts site performance while delivering better search results and drives better user engagement.

Future research can focus on grouping search results using Natural Language models integrated with behavior-based approaches. Continuously utilizing user feedback in selecting eligible queries with Reinforcement Learning can also be an exciting approach for further exploration. From a business perspective, the framework's utility to integrate curated experiences for users for specific high volume queries can also be explored.

\bibliographystyle{unsrt}
\bibliography{sample-sigconf}

\end{document}